\documentclass[%
preprint,
nofootinbib,
 amsmath,amssymb,
 aps,
]{revtex4-2}

\usepackage{graphicx}
\usepackage{dcolumn}
\usepackage{bm}
\usepackage{stmaryrd}

\newcommand{\Slash}[1]{{\ooalign{\hfil/\hfil\crcr$#1$}}}

\begin{document}


\title{Proton mass decompositions in the NNLO QCD}

\author{Kazuhiro Tanaka}
 \email{kztanaka@juntendo.ac.jp}
\affiliation{%
 Department of Physics, Juntendo University, Inzai, Chiba 270-1606, Japan
}%




\date{\today}

\begin{abstract}
Proton matrix elements of the QCD energy-momentum tensor (EMT) are expressed by the gravitational form factors.
The forward 
values of the gravitational form factors allow for a decomposition of the proton mass into contributions from quarks and gluons, and further subdivisions into contributions from quark masses and from the QCD trace anomaly may be considered. We present the most recent evaluations of these mass decompositions,
using a recent quantitative evaluation of the forward values of relevant gravitational form factors at the next-to-next-to-leading order (NNLO) QCD. We also calculate the renormalization scale dependence of each component within these decompositions.  
Furthermore, similar calculations are performed with another decomposition of the proton mass, 
organized strictly according 
to
the separation into the traceless 
part and trace 
part for each of the gauge-invariant quark/gluon parts of the EMT,
such that the former (twist-two) 
quark/gluon 
contributions of the EMT embody 
the effects of the partonic motions 
inside the proton,
while the latter (twist-four) contributions are induced as parton correlations by non-perturbative QCD interactions.
We demonstrate the advantages of this new decomposition.
We also present the results for the pion, which exhibit quite different parton-correlation behaviors from the proton.
\end{abstract}

\maketitle

\newpage
\section{Intoduction}
\label{sec1}

The energy-momentum tensor (EMT) plays fundamental roles in quantum field theories, arising in the generators for
the Poincar\'{e},
dilatation, and conformal transformations~\cite{Treiman:1986ep,Braun:2003rp}. In QCD, the EMT is composed of several composite operators,
and their physical interpretations are not straightforward due to their sophisticated renormalization properties~\cite{Collins:1984xc,Ji:1995sv,Hatta:2018sqd} and the associated trace anomalies~\cite{Nielsen:1977sy,Adler:1976zt,Collins:1976yq,Tanaka:2018nae,Ahmed:2022adh}. These complications, in particular, have been preventing attempts~\cite{Ji:1994av,Ji:2021mtz,Lorce:2017xzd,Metz:2020vxd,Ji:2021qgo,Lorce:2021xku,Ji:2025qax,Liu:2021gco,Liu:2023cse} to understand the mass structure of the proton from the EMT.
 Promoting such understanding is a hot topic because the origin of the proton mass is one of the major goals for the upcoming electron-ion collider (EIC)~\cite{Accardi:2012qut,AbdulKhalek:2021gbh}.

The symmetric EMT in QCD, $T^{\alpha\beta}
=  T^{\alpha\beta}_q+ T^{\alpha\beta}_g$, reads~\cite{Ji:1994av,Braun:2003rp,Polyakov:2018zvc}
\begin{align}
T^{\alpha\beta}_q= i\bar{\psi}\gamma^{(\alpha}\overleftrightarrow{D}^{\beta)}\psi ,\;\;\;
T_g^{\alpha\beta}= -F^{\alpha\lambda}F^\beta_{\ \lambda} + \frac{\eta^{\alpha\beta}}{4}F^2 ,
\label{tg}
\end{align}
with $T^{\alpha\beta}_q$ and $T_g^{\alpha\beta}$ being the gauge-invariant quark part and gluon part
of the bare EMT operator $T^{\alpha\beta}$, 
where
$\overleftrightarrow{D}^\alpha \equiv (\overrightarrow{D}^\alpha -\overleftarrow{D}^\alpha)/2$, $R^{(\alpha}S^{\beta)}\equiv (R^\alpha S^\beta+R^\beta S^\alpha)/2$, $F^2 \equiv F^{\alpha\lambda}F_{\alpha \lambda}$, $D^\alpha$ is the covariant derivative,
$F^{\alpha\lambda}$ is the gluon field strength tensor, and $\eta^{\alpha\beta}$ is the
metric tensor;
we have neglected the gauge-variant terms, i.e., the ghost terms and the gauge fixing term, as they do not affect our final results\footnote{The hadron matrix elements of those gauge-variant terms vanish; see e.g., \cite{Collins:1984xc,Kodaira:1998jn}.}.
Here and below, we assume the use of the QCD equations of motion\footnote{The hadron matrix elements of composite operators proportional to the equations of motion vanish~\cite{Collins:1984xc,Kodaira:1998jn}.}.
$T^{\alpha\beta}$ can be constructed from the canonical EMT by adding appropriate total-divergence contributions to make it symmetric, $T^{\alpha\beta}=T^{\beta\alpha}$, while preserving the continuity equation, 
\begin{align}
\partial_\alpha T^{\alpha\beta}=0.
\label{barecons}
\end{align}

For a (local) bare composite operator $O(x)$ at the space-time point $x^\alpha$, like those arising in (\ref{tg}), 
the corresponding  
renormalized composite operator in the MS-like (MS, $\overline{\rm MS}$)
schemes in the dimensional regularization in the $d$ space-time dimensions~\cite{Collins:1984xc}
shall be denoted\footnote{$\left[O(x)\right]$ corresponds to $\left(O(x)\right)_R$ in \cite{Hatta:2018sqd,Tanaka:2018nae}.} as $\left[O(x)\right]$; here, $\left[O(x)\right]$ depends on the renormalization scale $\mu$ in the MS-like schemes, although we do not show its dependence explicitly:
e.g., 
in the MS scheme, 
$\left[i\bar{\psi}\gamma^{(\alpha}\overleftrightarrow{D}^{\beta)}\psi\right]$
represents the corresponding bare operator
$i\bar{\psi}\gamma^{(\alpha}\overleftrightarrow{D}^{\beta)}\psi$
with the finite-part operation 
(i.e., the subtraction of the UV-divergent $\frac{1}{4-d}$ poles)
being performed,
where, due to
renormalization mixing, the corresponding UV poles are associated not only with $i\bar{\psi}\gamma^{(\alpha}\overleftrightarrow{D}^{\beta)}\psi$ but also with the other symmetric tensors, $F^{\alpha\lambda}F^\beta_{\ \lambda}$, $\eta^{\alpha\beta} F^2$, and $\eta^{\alpha\beta}m\bar{\psi}\psi$, as the gauge-invariant bare operators ($m$ denotes the bare quark mass)~\cite{Hatta:2018sqd,Tanaka:2018nae}.
Similarly, those four bare composite operators, 
\begin{align}
\left\{i\bar{\psi}\gamma^{(\alpha}\overleftrightarrow{D}^{\beta)}\psi, F^{\alpha\lambda}F^\beta_{\ \lambda}, \eta^{\alpha\beta} F^2, \eta^{\alpha\beta}m\bar{\psi}\psi\right\},
\label{opbasis}
\end{align}
mix under renormalization to calculate
$\left[F^{\alpha\lambda}F^\beta_{\ \lambda}\right]$. On the other hand, only $F^2$ and $m\bar{\psi}\psi$ participate in the renormalization mixing to calculate $\left[ F^2\right]$, while $m\bar{\psi}\psi$ is not renormalized, $m\bar{\psi}\psi=\left[ m\bar{\psi}\psi\right]$, see \cite{Tarrach:1981bi}; in such MS-like
schemes, it is straightforward to see that
the relations,
\begin{align}
\left[ T^{\alpha\beta}\right] =T^{\alpha\beta}, \;\;\; \partial_\alpha \left[ T^{\alpha\beta}\right]=0,
\label{renomcontinue}
\end{align}
are satisfied with 
\begin{align}
    \left[ T^{\alpha\beta}\right]=\left[ T_q^{\alpha\beta}\right]+\left[ T_g^{\alpha\beta}\right],
\label{totaltren}
\end{align}
and thus $\left[ T^{\alpha\beta}\right]$ does not receive the anomalous dimension; as is well-known, however, such MS-like
schemes lead to $\eta_{\alpha\beta}\left[i\bar{\psi}\gamma^{(\alpha}\overleftrightarrow{D}^{\beta)}\psi\right]\neq \left[i\bar{\psi}\overleftrightarrow{\Slash{D}}\psi\right]$, $\eta_{\alpha\beta}\left[ F^{\alpha\lambda}F^\beta_{\ \lambda}\right]\neq \left[ F^2\right]$. This is because ``taking a trace with $\eta_{\alpha\beta}$'' and ``the finite-part operation via the UV pole subtractions'' do not commute, resulting in the anomalies as in (\ref{tqrenaq}) below; by contrast, for the bare operators, $\eta_{\alpha\beta}i\bar{\psi}\gamma^{(\alpha}\overleftrightarrow{D}^{\beta)}\psi=i\bar{\psi}\overleftrightarrow{\Slash{D}}\psi=m\bar{\psi}\psi$, as well as $\eta_{\alpha\beta}F^{\alpha\lambda}F^\beta_{\ \lambda}=F^2$, holds.

One might consider the possibility of adopting $\eta_{\alpha\beta}\frac{4}{d}F^2, \eta^{\alpha\beta}\frac{4}{d}m\bar{\psi}\psi$ instead of $\eta_{\alpha\beta}F^2, \eta^{\alpha\beta}m\bar{\psi}\psi$ in the  operator basis for renormalization in the $d$ space-time dimensions.
Then, the finite-part operation by the subtractions of the UV poles associated with the basis of the independent operators, $\left\{i\bar{\psi}\gamma^{(\alpha}\overleftrightarrow{D}^{\beta)}\psi, F^{\alpha\lambda}F^\beta_{\ \lambda}, \eta^{\alpha\beta} \frac{4}{d}F^2, \eta^{\alpha\beta}\frac{4}{d}m\bar{\psi}\psi\right\}$,
gives
a particular renormalization scheme; 
for a bare operator $O(x)$, we denote the corresponding renormalized operator in this particular scheme as $\left \llbracket O(x)\right\rrbracket$.
This particular scheme is known~\cite{Collins:1974da} to allow us to commute ``taking a trace with $\eta_{\alpha\beta}$'' and ``the finite-part operation via the UV pole subtractions''; indeed, e.g., by explicit one-loop calculation, one can show,
$\eta_{\alpha\beta}\left \llbracket i\bar{\psi}\gamma^{(\alpha}\overleftrightarrow{D}^{\beta)}\psi\right\rrbracket=\left \llbracket \frac{4}{d}m\bar{\psi}\psi\right\rrbracket=\left \llbracket m\bar{\psi}
\psi\right\rrbracket
=\left \llbracket
i\bar{\psi}\overleftrightarrow{\Slash{D}}\psi\right\rrbracket$, and
$\eta_{\alpha\beta}\left \llbracket F^{\alpha\lambda}F^\beta_{\ \lambda}\right\rrbracket
=\left \llbracket \frac{4}{d}F^2\right\rrbracket
+\left \llbracket \left( \frac{
d-4}{4}\right)\frac{4}{d}F^2\right\rrbracket=\left \llbracket F^2\right\rrbracket$, 
which demonstrate the corresponding commutativity. 
Here,
the trace anomaly actually arises in the term $\left \llbracket \left( \frac{
d-4}{4}\right)\frac{4}{d}F^2\right\rrbracket$.
However, this particular scheme has the disadvantage:
the formula,
$T_g^{\alpha\beta}= -F^{\alpha\lambda}F^\beta_{\ \lambda} + \frac{d}{16}\times \eta^{\alpha\beta}\frac{4}{d}F^2$, from (\ref{tg}) implies $\left \llbracket
T^{\alpha\beta}\right\rrbracket\neq
T^{\alpha\beta}$, with $\left \llbracket
T^{\alpha\beta}\right\rrbracket=\left \llbracket
T_q^{\alpha\beta}\right\rrbracket+\left \llbracket
T_g^{\alpha\beta}\right\rrbracket$,
because the finite-part operation by the corresponding UV-pole subtractions 
is not a linear operation
for the $d$-dependent coefficient $\frac{d}{16}$ in front of $\eta^{\alpha\beta}\frac{4}{d}F^2$
in $T_g^{\alpha\beta}$, and thus gives rise to the additional anomalous contribution.
As a result, $\left \llbracket
T^{\alpha\beta}\right\rrbracket$ is not conserved, as~\cite{Collins:1984xc}
\begin{align}
    \partial_\alpha \left \llbracket T^{\alpha\beta}\right\rrbracket\neq 0,
\label{noncons}
\end{align}
leading to nonzero anomalous dimensions for $\left \llbracket
T^{\alpha\beta}\right\rrbracket$.

To derive the QCD formulas for the mass structure of the proton from the EMT, it is essential to ensure the properties~(\ref{renomcontinue}).
In this paper, we assume the MS-like schemes by subtracting the UV poles associated with the operator basis~(\ref{opbasis}),
thereby (\ref{renomcontinue}) is ensured.
The renormalization of the QCD EMT~(\ref{tg}) in such MS-like schemes is worked out with the accuracy of one- as well as two-loop~\cite{Hatta:2018sqd}, three-loop~\cite{Tanaka:2018nae}, and four-loop~\cite{Ahmed:2022adh}, and these results also demonstrate that each of $\left[ T_q^{\alpha\beta}\right]$ and $\left[ T_g^{\alpha\beta}\right]$ in (\ref{totaltren}) receives a definite amount of trace anomalies, whose total sum reproduces the well-known QCD trace anomaly of \cite{Collins:1976yq,Nielsen:1977sy}.
These results should be useful for deriving the proton matrix elements of the traceless part as well as the trace part from each of $\left[ T_q^{\alpha\beta}\right]$ and $\left[ T_g^{\alpha\beta}\right]$ in the EMT,
systematically taking into account the higher order corrections, and therefore allowing us to investigate the
quark and gluon contributions to the proton mass structure in a model-independent formulation.

The values of the proton matrix elements of each part of the EMT are encoded in the corresponding gravitational form factors~\cite{Polyakov:2018zvc,Tanaka:2018wea}. 
In addition to the gravitational form factors, there are various form factors that reflect the internal structures of the proton, and those form factors are generally complicated non-perturbative quantities. 
Remarkably, however, it has been pointed out~\cite{Hatta:2018sqd,Tanaka:2018nae} that the forward (zero momentum transfer) gravitational form
factors, 
relevant 
for evaluations of the proton mass structure, 
get simplified owing to the QCD constraints from (anomalous) conservation laws, manifest for an expectation value with a proton state; i.e., momentum conservation due to (\ref{renomcontinue}) and the broken scale invariance due to quantum anomalies.
As a result, the forward gravitational form
factors
have been proven to be disentangled into a few non-perturbative parameters multiplied by certain renormalization-group-improved perturbative contributions,
and, furthermore, the corresponding non-perturbative parameters can be determined by empirical information.
With such a model-independent framework, 
a quantitative evaluation
of the forward values of the quark and gluon contributions to the proton gravitational form factors 
has recently been performed~\cite{Tanaka:2022wzy}
at the next-to-next-to-leading order
(NNLO) QCD, taking into account the perturbative effects up to the three-loop order~\cite{Tanaka:2018nae}, and the results provide us with the forward values of the relevant gravitational form factors at the accuracy of a few percent level, for the  various renormalization scales as controlled by the three-loop renormalization group (RG)
equations.

Utilizing those NNLO results of the gravitational form factors of \cite{Tanaka:2022wzy}, this paper presents the first evaluations of proton mass decompositions in the NNLO QCD,
providing us with a model-independent result
on the proton mass structure with the ever highest accuracy. We present our NNLO results not only at a typical hadron scale $\mu=2$~GeV but also at the other renormalization scales $0.7~{\rm GeV} \lesssim \mu \lesssim 20~{\rm GeV}$, as governed by the three-loop RG
equations.

First of all, we apply our framework to the proton mass decomposition formulas derived in previous works~\cite{Ji:1994av,Ji:2021mtz,Lorce:2017xzd,Metz:2020vxd,Ji:2021qgo,Lorce:2021xku,Ji:2025qax}, and present the NNLO update of the evaluations of those decompositions.
We show the corresponding NNLO results over a range of the renormalization scales $0.7~{\rm GeV} \lesssim \mu \lesssim 20~{\rm GeV}$, and explain their characteristic behaviors and some issues relating to the fact that the previous decompositions~\cite{Ji:1994av,Ji:2021mtz,Lorce:2017xzd,Metz:2020vxd,Ji:2021qgo,Lorce:2021xku,Ji:2025qax} resulted in ``incomplete separation'' into the traceless part and trace part of the EMT.
This fact suggests another decomposition
which is organized strictly according to the separation
into the traceless part and trace part for each of the gauge-invariant quark part and gluon part of the
EMT in the MS-like schemes\footnote{A formally similar separation strictly into the traceless part and the trace part for the proton mass
has been discussed in \cite{Liu:2021gco,Liu:2023cse},
where the trace part is subdivided by a scheme different from ours. 
In \cite{Liu:2021gco,Liu:2023cse}, the corresponding proton mass decomposition is presented at one hadron scale $\mu =2$~GeV, as well as at one weak scale $\mu=250$~GeV.}. We demonstrate that this new decomposition is free from those issues and corresponds to the physical picture that the traceless (i.e., twist-two) quark/gluon parts of the EMT embody the effects
of the partonic motions inside the proton, while the trace (i.e., twist-four) contributions are induced
as parton correlations~\cite{Jaffe:1983hp,Shuryak:1981kj,Kodaira:1998jn} by non-perturbative QCD interactions.

We also apply our NNLO framework to the pion case to calculate the 
pion-mass decompositions:
we are able to update the 
pion-mass decomposition results in the literature and also extend the calculation to that with the above-mentioned ``strict twist-decomposition'',
which proves to be useful for the pion case,
similarly to the proton case. Moreover, comparing the results between the pion and proton cases indicates that 
the effects of the partonic motions are similar for both cases, while the parton-correlation effects from non-perturbative QCD interactions show quite different behaviors. Mathematically strict twist-decomposition allows such characterization of the physical effects inside hadrons, as the twist-two effects for the former and the twist-four effects for the latter.

The rest of the paper is organized as follows: in Sec.~\ref{sec2}, the proton mass is expressed as the proton matrix element of the EMT of (\ref{totaltren}), as the rest energy as well as the invariant mass,
and its decompositions into the various QCD effects discussed in the literature are explained.
The terms of those decompositions are
represented by the gravitational form factors that parameterize the proton matrix elements of the EMT, and the recent NNLO evaluation of the relevant
gravitational form factors is explained; combining them, the NNLO update of those decompositions is presented. Their renormalization scale dependence is also discussed. 
In Sec.~\ref{sec3}, similar calculations are performed with the rearranged proton-mass decomposition such that the traceless part and trace part are strictly separated 
for each of the gauge-invariant quark/gluon parts of the EMT in the MS-like schemes. We demonstrate that this new decomposition improves the issues 
that arise in the calculations in Sec.~\ref{sec2}. Those frameworks are applied to the pion case in Sec.~\ref{sec4}, and the comparison with the proton case is discussed, revealing the physical effects inside the hadrons implied by our new decomposition. Sec.~\ref{sec5} is reserved for conclusions.

\section{Update of mass decompositions from  gravitational form factors at NNLO}
\label{sec2}

We use the EMT expressed by the composite operators renormalized in the MS-like schemes, so that (\ref{renomcontinue}) is satisfied.
Our proton state $|p\rangle$ has the eigenvalue $p^\alpha$
for the 
4-momentum operator defined as 
the volume integral of $\left[T^{0 \alpha}\right]$ and is normalized as $\langle p'|p\rangle =2p^0 (2\pi)^3 \delta^{(3)}(p'-p)$.
Then, translation invariance allows us to show
$\langle p|\left[T^{\alpha\beta}\right]|p\rangle 
=2p^\alpha p^\beta$, from which the well-known expressions for the proton mass $M$ read,
\begin{align}
2 M^2 =\left.  \langle p|\left[T^{00}\right]|p\rangle\right|_{\bm{p}=\bm{0}},
\;\;\;\;\;\;
2 M^2 = \langle p|\eta_{\alpha \beta} \left[T^{\alpha \beta}\right]|p\rangle  ,
\label{m22}
\end{align}
as the ``rest energy'' and as the ``invariant mass'', respectively\footnote{With (\ref{renomcontinue}) in the MS-like schemes used in this paper, 
$\eta_{\alpha \beta} \left[T^{\alpha \beta}\right]= \left[\eta_{\alpha \beta}T^{\alpha \beta}\right]$, i.e., ``taking trace'' and ``the finite-part operation'' commute for the total EMT, but note, 
$\eta_{\alpha \beta} \left[T_r^{\alpha \beta}\right]\neq \left[\eta_{\alpha \beta}T_r^{\alpha \beta}\right]$,
for $r=q,g$.}.

The invariant mass of (\ref{m22})
leads to the well-known proton mass formula~\cite{Shifman:1978zn}
associated with the
trace 
anomaly~\cite{Nielsen:1977sy,Adler:1976zt,Collins:1976yq}, i.e., 
\begin{align}
\eta_{\alpha \beta} \left[T^{\alpha \beta}\right]=\left[T^\alpha_{\ \alpha} \right]
=\frac{\beta(g)}{2g} \left[F^2\right]
+\left(1+\gamma_m(g)\right)\left[m \bar{\psi}\psi\right]
\ , 
\label{totalanomaly}
\end{align}
with
the QCD $\beta$-function $\beta(g)$ and
the anomalous dimension for the quark mass, $\gamma_m(g)$,
as functions of the renormalized QCD coupling constant $g$.
Thus, the proton mass is expressed by the non-perturbative matrix elements $\langle p|\left[ F^2\right]|p\rangle$ and
$\langle p|\left[m \bar{\psi}\psi\right]|p\rangle$; these matrix elements of the Lorentz scalar operators equal the corresponding matrix elements in the rest frame, as $\langle p|\left[ O(0)\right]|p\rangle$
$= \langle p|\left[ O(0)\right]|p\rangle|_{\bm{p}=\bm{0}}
= 2M\langle \int d^3 x\left[ O(x)\right]  \rangle$, with $O(x)=F^2(x), m \bar{\psi}(x)\psi(x)$, 
where we have introduced the shorthand notation for the {\it expectation value in the rest frame} as
\begin{align}
\langle\, \cdots\,  \rangle\equiv  \left. \frac{\langle p|\cdots|p\rangle}{\langle p| p\rangle}\right|_{\bm{p}=\bm{0}}.
\label{shorthandshorthand}
\end{align}
As will be discussed below, $\langle p|\left[ m \bar{\psi}\psi\right]|p\rangle$ is expressed by the sigma terms determined (semi)empirically; then, the condition that the second equation of (\ref{m22}) using (\ref{totalanomaly}) should reproduce the value of the proton mass allows us to evaluate the proton matrix element $\langle p|\left[ F^2\right]|p\rangle$.
Now, using those non-perturbative values of $\langle p|\left[ F^2\right]|p\rangle$,
as well as of
$\langle p|\left[ m \bar{\psi}\psi\right]|p\rangle$, the separation of $\left[ T^{\alpha\beta}\right]$
into (\ref{totaltren})
allows us to decompose the proton mass into the quark and gluon contributions as~\cite{Tanaka:2018nae}
\begin{align}
M={\cal M}_q+{\cal M}_g, \;\;\;\;\;\;\;\;\;\;\;\;\;\;\;
{\cal M}_r=\frac{\langle p| \eta_{\alpha\beta}\left[T_r^{\alpha\beta}\right] |p\rangle}{2M} \;\;\;\;\;\; (r=q, g),
\label{calmqg0}
\end{align}
in terms of the trace anomaly for each term in the RHS of (\ref{totaltren}), {\it as ``taking trace with $\eta_{\alpha\beta}$'' after  ``the finite-part operation via the UV pole subtractions''}~\footnote{
If their order were reversed, as taking trace before the finite-part operation, the RHS of (\ref{tqrenaq}) would be modified into
$m \bar{\psi}\psi$ as mentioned below (\ref{totaltren}),
which corresponds to the ``D2 scheme''\cite{Metz:2020vxd,Lorce:2021xku,Liu:2021gco}.}:
the trace anomaly for the quark part reads~\cite{Hatta:2018sqd}
\begin{align}
\eta_{\alpha \beta} \left[ T_q^{\alpha \beta}\right]=
\left[  m \bar{\psi}\psi\right]+ \frac{{{\alpha _s}}}{{4\pi }}\left(\frac{{{n_f}}}{3}\left[F^2\right]+
\frac{{4{C_F}}}{3}\left[m \bar{\psi}\psi\right]\right)+\cdots ,
\label{tqrenaq}
\end{align}
with $\alpha_s\equiv g^2/4\pi$, for $n_f$ flavor and
$N_c$ color with 
$C_F=(N_c^2-1)/2N_c$.
The formula for the gluon part $\eta_{\alpha \beta} \left[T_g^{\alpha \beta}\right]$ is also obtained similarly,
and coincides with $\eta_{\alpha \beta} \left[T^{\alpha \beta}\right]- \eta_{\alpha \beta} \left[T_q^{\alpha \beta}\right]$, where we have (\ref{totalanomaly}). The formula~(\ref{tqrenaq}) is derived in the MS-like 
schemes, and the ellipses 
stand for the corrections at the two loops (${\cal O}(\alpha_s^2)$) and higher, 
which are derived in \cite{Hatta:2018sqd,Tanaka:2018nae,Ahmed:2022adh}.
The quantitative results of (\ref{calmqg0}) using the three-loop 
formulas~\cite{Tanaka:2018nae} for $\eta_{\alpha \beta} \left[ T_q^{\alpha \beta}\right]$ and $\eta_{\alpha \beta} \left[T_g^{\alpha \beta}\right]$  will be discussed in Figs.~\ref{fig2}, \ref{fig4} in 
Secs.~\ref{sec3},~ \ref{sec4} below.

The rest energy of (\ref{m22}) reads
$M=(\langle p|\left[ T^{00}\right]|p\rangle/2M)_{\bm{p}=\bm{0}} 
$$=\langle\, H\,  \rangle$
using (\ref{shorthandshorthand}),
where 
\begin{align}
H=\int d^3 x \bigl(\left[ T_q^{00}(x)\right]+\left[T_g^{00}(x)\right]\bigr)
=
\int d^3 x \Bigl\{\left[\bar{\psi}\bigl( i\bm{\gamma}\cdot \overleftrightarrow{\bm{D}} 
+m
\bigr)\psi\right]+\frac{1}{2}\bigl(  \left[\bm{E}^2\right] +\left[\bm{B}^2\right]\bigr)\Bigr\},
\label{hamiltonian}
\end{align}
denotes the QCD Hamiltonian composed of the quark Dirac Hamiltonian and the chromoelectric/chromomagnetic Hamiltonian,
as derived from~(\ref{totaltren}), (\ref{tg}) using the QCD equations of motion,
so that $M=\langle\, H\,  \rangle$ leads to the mass formula as the sum of the rest energies, $M=U_q + U_g$, due to the quark and gluonic degrees of freedom.
The corresponding expectation values of the quark and gluon Hamiltonian operators of (\ref{hamiltonian}) 
could be computed by non-perturbative methods like lattice QCD~\cite{Yang:2018nqn}.
Here, alternatively, we represent and evaluate them using 
the gravitational form factors, $A_r (\Delta^2, \mu)$ and 
$\bar{C}_r(\Delta^2, \mu)$ with $r=q, g$,
which are the functions of $\Delta^2\equiv (p'-p)^2$ as well as the renormalization scale $\mu$, and are defined for each of the quark/gluon parts of (\ref{totaltren}) as
\begin{align}
\!\!\!
\langle p'| \left[T_r^{\alpha\beta} \right]|p\rangle 
= \bar{u}(p')\bigl[A_r (\Delta^2, \mu)
\gamma^{(\alpha}\overline p^{\beta)}
 + \bar{C}_r(\Delta^2, \mu)
 M\eta^{\alpha\beta}\bigr] u(p)+\cdots ,
\label{para}
\end{align}
where
$\overline p^\alpha\equiv (p^\alpha+p'^\alpha)/2$,
$u(p)$ is the proton spinor, and the ellipses stand for the terms associated with the other Lorentz structures that vanish as $\Delta=p'-p \to 0$ (see e.g., \cite{Polyakov:2018zvc,Tanaka:2018wea,Tong:2022zax,Fujii:2025aip}).
The sum rule, $\bar{C}_q(\Delta^2, \mu)+ \bar C_g(\Delta^2, \mu)=0$,
is derived
by contracting (\ref{para}) with $\Delta_\alpha$
and using
$\partial_\alpha\left[T^{\alpha\beta}\right]=0$ of  (\ref{renomcontinue}). Also, $\langle p|\left[ T^{\alpha\beta}\right]|p\rangle 
=2p^\alpha p^\beta$ ensures 
the forward sum rule, $A_q(0,\mu)+A_g(0,\mu)=1$.
Physically, the gravitational form factors represent the coupling of the proton with a graviton.
Instead of probing them via such impractical graviton-proton processes,
the gravitational form factors can be accessed through the spin-2 contributions induced by
the two vector currents; e.g., with the two electromagnetic currents, 
the deeply virtual Compton scatterings~\cite{Ji:1996ek} in JLab~\cite{Burkert:2018bqq} and the upcoming EIC~\cite{Accardi:2012qut,AbdulKhalek:2021gbh} allow us to probe the gravitational form factors. Substituting the forward limit of (\ref{para}) into 
$M=\langle\, H\,  \rangle$, 
we obtain the ``two-term decomposition''~\cite{Lorce:2017xzd}
using the $\Delta = 0$ values of the gravitational form factors as
\begin{align}
M=U_q + U_g , \;\;\;\;\;\;\;\;\;\;
U_r=M\left(A_r (0, \mu)
  + \bar{C}_r (0, \mu)\right).
\label{newuq}
\end{align}

Recently, the values of $\bar C_r(0, \mu)$, as well as of $A_r(0, \mu)$, as a function of $\mu$
have become available in
the NNLO, at the level of accuracy with the ${\rm uncertainty} \lesssim$ a few percent.
It is known~\cite{Polyakov:2018zvc,Tanaka:2018wea} that the matrix element~(\ref{para}) for $A_r(0, \mu)$ coincides with the moment of the proton's parton distribution
functions (PDFs) at the scale $\mu$ as
$A_q(0,\mu )= \sum_f \int_0^1dx x (q_f(x, \mu)+q_{\bar{f}}(x, \mu))$,
summed over the quark and antiquark distributions of flavor $f=u,d, \ldots$, 
and similarly for $A_g(0,\mu)$ using the gluon distribution.
The $\mu$ dependence of
$A_r(0, \mu)$
is controlled by the RG equations for the operators~(\ref{opbasis}), 
and it coincides with that of the first moment of the
evolution equations for the
PDFs. 
The explicit NNLO formulas of  $A_q(0, \mu)$ in the $\overline{\rm MS}$ scheme, as 
the solution of the corresponding three-loop RG 
equations, are given 
in e.g., \cite{Tanaka:2022wzy},
containing one single non-perturbative parameter as the value at a starting scale of the RG evolution; e.g., 
$A_q\left(0, \mu_i\right)=0.613$ at the starting scale $\mu_i \equiv 1.3~{\rm GeV}$ 
is determined in the $\overline{\rm MS}$ scheme
by the NNLO global QCD analysis of the experimental data for the nucleon's PDFs (CT18~\cite{Hou:2019efy}), representing the average total momentum fraction carried by the three flavors, $u, d,s$,
which are the active quark flavors at the starting scale $\mu_i$.
The uncertainties for $A_r(0, \mu)$ as the NNLO solution are of a few percent level.
For $\mu_0\equiv 2$~GeV,  the CT18 global analysis with the variable
flavor number~\cite{Hou:2019efy} gives,
$\sum_{f=u,d,s} \int_0^1dx x [q^{\rm CT18}_f(x, \mu_0)+q^{\rm CT18}_{\bar{f}}(x, \mu_0)]=0.576(11)$, 
$A_g^{\rm CT18}(0, \mu_0)=0.414(8)$, 
while the above input value, $A_q(0, \mu_i)=0.613$,
is evolved to $\left. A_q(0, \mu_0)\right|_{{n_f} = 3}=0.574$, 
$\left. A_g(0, \mu_0)\right|_{{n_f} = 3}
=0.426$, in the NNLO with the fixed flavor number $n_f=3$; thus, in the following evaluations, we shall use, 
\begin{align}
\!\!\!\! \left. A_q\left(0, \mu_0\right)\right|_{{n_f} = 3}
=0.57(1) ,\ \;\;\;\;\;\;
\left. A_g\left(0, \mu_0\right)\right|_{{n_f} = 3}
=0.43(1) .
\label{aat2}
\end{align}

The values of $\bar C_r(0, \mu)$ are also available with the same level of accuracy as in $A_r(0, \mu)$:
the
trace of both sides of (\ref{para}) allows us to write down the formula~\cite{Hatta:2018sqd},
\begin{align}
\bar{C}_r(0, \mu)
=-\frac{1}{4}A_r (0, \mu)+\frac{1}{8M^2} \langle p| \eta_{\alpha\beta}\left[T_r^{\alpha\beta}\right]|p\rangle  
\;\;\;\;\;\;\;\;(r=q, g),
\label{bytw4byac}
\end{align}
and both terms in the RHS are evaluated at the NNLO accuracy~\cite{Tanaka:2022wzy}, substituting
the above-mentioned three-loop evaluation of $A_r(0, \mu)$ and of (\ref{tqrenaq}), (\ref{totalanomaly}).
The explicit NNLO formula~\cite{Tanaka:2022wzy} of $\bar C_r(0, \mu)$ in the $\overline{\rm MS}$ scheme
gives\footnote{We note that ${\bar C_q(0, \mu_0 )} |_{n_f = 3}=-0.163(3)$ presented in \cite{Tanaka:2022wzy}
shows the uncertainty due to those of the sigma terms $\sigma_{\pi N}, \sigma_s$, while (\ref{bytw4byac}) tells us that it should be combined with the uncertainty of (\ref{aat2}).
This results in~(\ref{explicitresults}).}
\begin{align}
\left. {\bar C_q(0, \mu_0 )} \right|_{{n_f} = 3}=-\left. {\bar C_g(0, \mu_0 )} \right|_{{n_f} = 3}=-0.163(4) .
\label{explicitresults}
\end{align}
Those NNLO 
values~(\ref{aat2}), (\ref{explicitresults}) 
give 
$M  =[0.41(1)+	0.59(1)]M$
at the scale $\mu=\mu_0$ ($= 2$~GeV), using (\ref{newuq}).

This result may be elaborated by separating the contribution due to the quark mass term in the Dirac Hamiltonian of (\ref{hamiltonian}), so that the ``three-term decomposition''
is obtained as
$M=(U_q-M_m)+U_g
+M_m$~\cite{Metz:2020vxd,Lorce:2021xku}, with
\begin{align}
M_m\equiv \bigl\langle\,  \int d^3 x  \left[m\bar{\psi}
\psi\right]\,   \bigr\rangle
 =
 \frac{
\langle p|\left[m \bar{\psi}
\psi \right]|p\rangle
}{2M}.  
\label{m3q}
\end{align}
Guided by the usual manipulation for a (renormalized) second rank tensor to separate its traceless and trace parts as, 
$\left[T_r^{\alpha\beta}\right]=\left(\left[T_r^{\alpha\beta}\right]-\eta^{\alpha\beta}\frac{1}{4}\eta_{\lambda\rho}\left[T_r^{\lambda\rho}\right]\right)+\eta^{\alpha\beta}\frac{1}{4}\eta_{\lambda\rho}\left[T_r^{\lambda\rho}\right]$, 
the contributions corresponding to the trace anomaly can be extracted from $U_q$ and $U_g$,
so that the three-term decomposition may be further reorganized as
\begin{align}
M=M_q+M_g +M_m+M_a,
  \label{4term}
\end{align}
where
$M_q=  U_{q}-M_m-
  \frac{1}{4}({\cal M}_q-M_m)$, 
$M_g=U_g-\frac{1}{4}{\cal M}_g$
(see (\ref{calmqg0}), (\ref{tqrenaq})), and  
\begin{align}
M_a \equiv \frac{1}{4}({\cal M}_q-M_m+{\cal M}_g)
=\frac{1}{4}\bigl\langle\  \int d^3 x
\bigl(\ \frac{\beta(g)}{2g} \left[F^2\right]
+\gamma_m(g)\left[m \bar{\psi}\psi\right]\
\bigr)\
  \bigr\rangle
\label{ma}
\end{align}
corresponds to the total trace-anomaly
contribution in (\ref{totalanomaly}).
The formula~(\ref{4term}) corresponds to the famous ``four-term decomposition'' proposed by Ji~\cite{Ji:1994av,Ji:1995sv,Ji:2021mtz}
and coincides with Eq.(62) of \cite{Lorce:2021xku}. 
For the $n_f=3$ active quark flavors, 
$M_m$ of (\ref{m3q}) can be expressed as
$M_m
=\frac{1}{2M}
 \langle p |\left[\frac{m_u+m_d}{2} (\bar u u+\bar d d)\right]+\left[m_s \bar s s\right] | p\rangle
 =\sigma_{\pi N}+ \sigma_{s}$, with
the pion-nucleon sigma-term $\sigma_{\pi N}$ and the strangeness content $\sigma_{s}$, 
up to small isospin-violating corrections of ${\cal O}(m_d-m_u )$.
Here, we use $\sigma_{\pi N}=0.0591(35)$~GeV
due to a recent phenomenological analysis~\cite{Hoferichter:2015dsa} 
and $\sigma_{s}=0.0456(62)$~GeV
due to a recent lattice QCD determination~\cite{Alexandrou:2019brg};
see also \cite{Alarcon:2011zs,RuizdeElvira:2017stg,Hoferichter:2015hva,Alarcon:2012nr,Junnarkar:2013ac,Ren:2012aj,Ren:2014vea,Ren:2017fbv,Gupta:2021ahb,Yang:2015uis,Yamanaka:2018uud,XQCD:2013odc,FlavourLatticeAveragingGroupFLAG:2024oxs}.
These non-perturbative matrix elements arise also in 
(\ref{bytw4byac}) with (\ref{tqrenaq}), (\ref{totalanomaly}),
and the above-mentioned NNLO values of $\bar C_{q,g}(0, \mu)$ (e.g., (\ref{explicitresults}))
are calculated using the same input values
for $\sigma_{\pi N}, \sigma_{s}$.

\begin{figure}[hbtp]
\centering
\hfill
\begin{minipage}{0.35\columnwidth}
\centering
\includegraphics[width=\columnwidth]{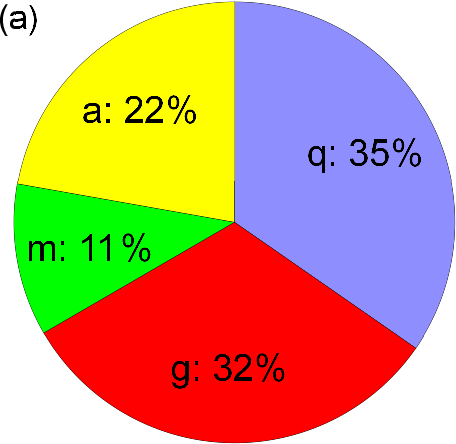}
\vspace{0.5cm}
\end{minipage}
\hfill
\begin{minipage}{0.6\columnwidth}
\centering
\includegraphics[width=\columnwidth]{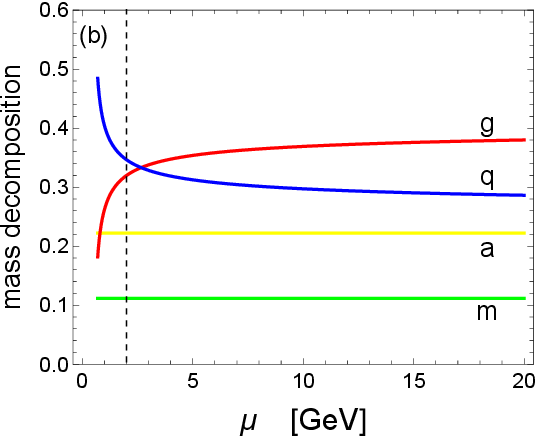}
\end{minipage}
\caption{The ratios
$M_j/M$ ($j=q, g, m, a$) in Ji's four-term decomposition of the proton mass, (\ref{4term}), in the NNLO QCD with the $n_f=3$ active quark flavors and the $\overline{\rm MS}$ scheme:
(a) at $\mu=2$~GeV; (b) as functions of $\mu$ with the dashed line indicating  
the case (a).
}
\label{piechart}
\end{figure}

We now calculate each term of (\ref{4term})
(see also (\ref{mqmgexplicit}) below)
using the above-explained non-perturbative inputs, and this 
provides the NNLO update for the evaluation of Ji's four-term decomposition of the proton mass, 
as shown in Fig.~\ref{piechart}(a) for a typical hadron scale $\mu=2$~GeV.
Our numerical calculations are performed for the fixed number $n_f=3$ active quark flavors
using the three-loop $\overline{\rm MS}$ running coupling for $\alpha _s(\mu )$~\footnote{
We determine $\alpha _s(\mu )$ 
as the exact numerical solution~\cite{Herren:2017osy} of  
the corresponding RG equation
controlled by the three-loop $\beta$ function. We use $\Lambda_{\rm QCD}^{(3)}\simeq0.3359$~GeV,
i.e.,
$\alpha_s(\mu=1~\rm{GeV})\simeq 0.4736$, as in \cite{Tanaka:2022wzy}.}.
Fig.~\ref{piechart}(b) shows the $\mu$ dependences of each of the four terms
in (\ref{4term}), i.e., 
the results of the NNLO QCD evolution from each contribution 
of Fig.~\ref{piechart}(a): $M_q$ and $M_g$ exhibit significant $\mu$ dependence with their crossing 
around the hadronic scale, while $M_m$ as well as $M_a$ is RG invariant, see
(\ref{m3q}), (\ref{ma}).
Those NNLO results for the proton are accurate to a few percent uncertainties; e.g.,
the numbers presented inside the pie chart of Fig.~\ref{piechart}(a) have the uncertainties at most $\pm 1$ only in the last digit, and 
are consistent with the corresponding lattice QCD calculation~\cite{Yang:2018nqn} of $M=\langle\, H\,  \rangle$.
It is remarkable that the present NNLO QCD results, with a few (semi)empirical non-perturbative inputs, 
have uncertainties considerably smaller than those in the lattice QCD results.

We note that we encounter the following points (i)-(iii) in our evaluations of (\ref{4term}) for the proton\footnote{Similar points (i)-(iii) also arise in the three-term decomposition discussed above~(\ref{m3q}); e.g., (ii) was pointed out in \cite{Metz:2020vxd}.}: 

\noindent
(i) Strong $\mu$ dependence: one has to be careful regarding the strong $\mu$ dependence in Fig.~\ref{piechart}(b) when referring to the pie chart of Fig.~\ref{piechart}(a), as the latter 
would not represent a universal picture 
over the relevant scales.

\noindent
(ii) Potential heavy-quark contribution: the decoupling of the quarks heavier 
than the $s$-quark is not manifest for 
the quark mass term (\ref{m3q}); e.g., 
the inclusion of the charm content,
$\sigma_{c}=\frac{1}{2M}\langle p |\left[m_c \bar c c \right]
 | p 
 \rangle
=0.107(22)~{\rm GeV}$,
which is obtained by recent lattice QCD calculation~\cite{Alexandrou:2019brg} (see also \cite{XQCD:2013odc,FlavourLatticeAveragingGroupFLAG:2024oxs}),
would double the contribution of $M_m$, leading to its $\sim 100$\% uncertainty,
and also cause the corresponding modifications in $M_q$ as well as $M_a$.

\noindent
(iii) Incomplete separation into the traceless part and trace part: using the gravitational form factors of (\ref{para}), each term of (\ref{4term}) is expressed~\cite{Ji:1994av,Ji:1995sv} as
\begin{align}
M_q
=\frac{3}{4}MA_q (0, \mu)-\frac{3}{4}M_m,\;\;\;\;\;\;\;\;\;\;
M_g
=\frac{3}{4}MA_g (0, \mu),
\label{mqmgexplicit}
\end{align}
where $M_m$ of (\ref{m3q})
corresponds to the matrix element of the trace part (twist four). Similarly, $M_a$ of (\ref{ma}) corresponds to the trace part. 
Since $A_r(0, \mu)$ corresponds to the traceless part (twist two) as convinced shortly, $M_g$ corresponds to the traceless part, while $M_q$ is a mixture of the quantities corresponding to the traceless part and the trace part; therefore, (\ref{4term}) is not fully organized according to the separation into the traceless part and the trace part, i.e., according to the twist.

\section{Twist decomposition in the $\overline{\rm MS}$ scheme}
\label{sec3}

The composite operators are classified by the twist that equals their ``dimension minus spin'',
and 
the traceless part and the trace part of (\ref{totaltren}) are of twist two and four, respectively. Comparing this with~(\ref{para}),
$A_r(0,\mu)$ is of twist two, as mentioned in the point (iii) above.

The studies of hard processes like the deep inelastic lepton-proton scatterings clarified~\cite{Kodaira:1998jn} that  
the partonic motions inside the proton are described by matrix elements of the twist-two operators in QCD, 
while 
the non-perturbative QCD interactions 
induce correlations among the partons, emerging as
matrix elements of the operators of twist four and higher\footnote{For the quantities dependent on polarization, the twist-three operators also play roles in describing the spin-dependent partonic correlations; see e.g., \cite{Kodaira:1998jn,Balitsky:1996uh,Kodaira:1996md,Ball:1998sk}.}.
As noted in (iii), $M_q$ does not match with such physical classification according to the twist, although $M_q$ was 
intended~\cite{Ji:1994av,Ji:1995sv} to represent the contribution of the quark's gauge-invariant ``kinetic and potential energies'' operator $\left[\bar{\psi} i\bm{\gamma}\cdot \overleftrightarrow{\bm{D}} 
    \psi\right]$
of (\ref{hamiltonian}).
The physical meaning of the twist
calls for the rearrangement of the  decomposition~(\ref{4term}), such that
the terms of different twist in each of the quark and gluon contributions are completely separated:
the last term ($-\frac{3}{4}M_m$) of $M_q$ in (\ref{mqmgexplicit}), 
as well as $M_m$ of (\ref{4term}), is to be combined with $\frac{1}{4}({\cal M}_q-M_m)$ in (\ref{ma}), and this 
leads to another four-term decomposition given as
\begin{align}
\!
M=\frac{3}{4}MA_q (0, \mu)+\frac{3}{4}MA_g (0, \mu) +\frac{1}{4}{\cal M}_q+\frac{1}{4}{\cal M}_g,
\label{new4term}
\end{align}
where 
the last two terms of twist four are equal $\frac{1}{4}M$ in total, using (\ref{calmqg0}), but these terms are treated separately; this is because these terms correspond to the trace parts of the gauge-invariant quark operator and gluon operator, respectively,
expressed by the RHS of (\ref{calmqg0}), and read
\begin{align}
{\cal M}_r
=M\left(A_r (0, \mu)
  + 4\bar{C}_r(0, \mu)\right),
\label{calmqg}
\end{align}
using (\ref{bytw4byac}).
\begin{figure}[hbtp]
\centering
\hfill
\begin{minipage}{0.35\columnwidth}
\centering
\includegraphics[width=\columnwidth]{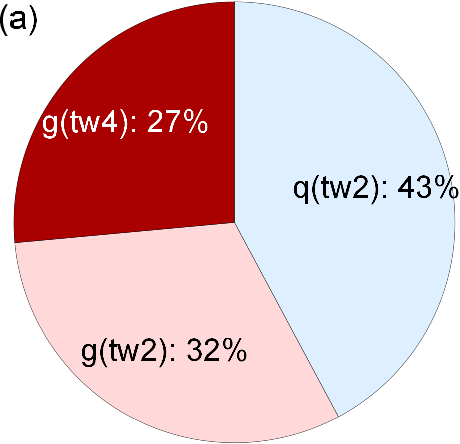}
\vspace{0.5cm}
\end{minipage}
\hfill
\begin{minipage}{0.6\columnwidth}
\centering
\includegraphics[width=\columnwidth]{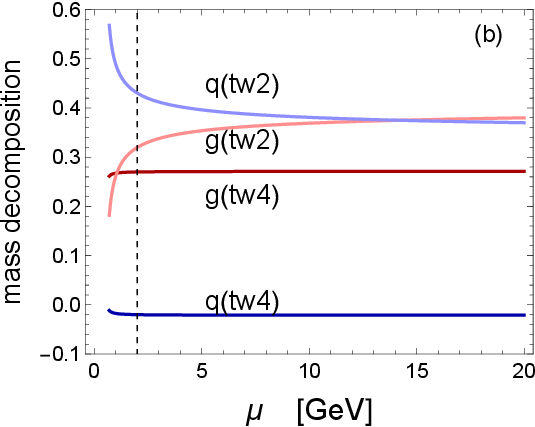}
\end{minipage}
\caption{Each term, divided by $M$, of another four-term decomposition of the proton mass, (\ref{new4term}), 
in the NNLO QCD  
with the $n_f=3$ active quark flavors and the $\overline{\rm MS}$ scheme:
(a) at $\mu=2$~GeV; (b) as functions of $\mu$ with the dashed line indicating  
the case (a).
}
\label{fig2}
\end{figure}
Remarkably, this allows us to resolve not only the point (iii) but also 
(i) and (ii), as we now demonstrate:
using the NNLO values (\ref{aat2}), (\ref{explicitresults}) and their evolution as functions of $\mu$, we evaluate each term of (\ref{new4term}) as displayed in Fig.~\ref{fig2}, 
which is accurate to a few percent uncertainties, similarly as in Fig.~\ref{piechart};
the numbers shown inside Fig.~\ref{fig2}(a) have the uncertainties at most $\pm 1$ only in the last digit.
Note that the twist-four quark contribution $\frac{1}{4}{\cal M}_q$ of (\ref{new4term}) is too small ($-2$\%~of $M$) to show up in Fig.~\ref{fig2}(a).
In Fig.~\ref{fig2}(b), we see that the $\mu$ dependence of the twist-two terms of~(\ref{new4term}) yields the similar values 
$\frac{3}{4}A_q(0, \mu)\simeq \frac{3}{4}A_g(0, \mu)\simeq 0.37$ 
for $10~{\rm GeV}\lesssim \mu\lesssim 20~{\rm GeV}$~\footnote{
The approach to the well-known asymptotic limit, $A_q(0, \infty)
  = \frac{n_f}{4C_F+n_f}$ and $A_g(0, \infty)
  = \frac{4C_F}{4C_F+n_f}$, is quite slow; see, e.g., \cite{Tanaka:2022wzy}.};
hence, the crossing of the quark and gluon twist-two contributions around the hadronic scale does not arise in Fig.~\ref{fig2}(b), and (i) is cleared. Now, Fig.~\ref{fig2}(a) could represent a universal qualitative picture of the proton-mass structure.

In our four-term decomposition~(\ref{new4term}),
the ${\cal O}(\alpha_s^0)$ contribution associated with the quark mass operator of (\ref{m3q}) 
is organized 
as the contribution $\frac{1}{4}M_m$ arising in the twist-four term $\frac{1}{4}{\cal M}_q$; see (\ref{calmqg0}), (\ref{tqrenaq}).
When  
$\sigma_{Q}=\frac{1}{2M}\langle p |\left[m_Q \bar Q Q \right]
 | p 
 \rangle$ of (ii) due to a heavy quark $Q$ is included by the replacements, $\frac{1}{4}M_m \to \frac{1}{4}(M_m+\sigma_Q)$ and $n_f \to n_f +1$, in this twist-four term $\frac{1}{4}{\cal M}_q$, 
the cancellation of the leading terms of
those new heavy-quark contributions is ensured by the heavy-quark
expansion of the corresponding mass operator,
\begin{equation}
  \langle h|\left[ m_Q\bar Q Q \right]|h \rangle=
-\frac{1}{12\pi}\langle h| \alpha_s\left[ F^2\right]  |h\rangle 
+ \cdots \ ,
\label{hqexp0}
\end{equation} 
which was derived in \cite{Shifman:1978zn,Shifman:1978bx,Generalis:1983hb}
for a state $|h \rangle$ with no valence
heavy quark $Q$, calculating the triangle-type heavy-$Q$ loop diagrams 
that induce the long-distance gluon fields, and the ellipses denote the corrections suppressed by two or more inverse-powers of 
$m_Q$.
Therefore, the decoupling of a heavy flavor $Q$ is manifest using (\ref{hqexp0}) for the combination of the operators in~(\ref{tqrenaq}) relevant to $\frac{1}{4}{\cal M}_q$
of (\ref{new4term}); this logic may be extended to higher orders with the help of the formulation of \cite{Chetyrkin:1997un}.  Now (ii) is resolved.

\section{The Pion case}
\label{sec4}

The treatments discussed in Secs.~\ref{sec2} and \ref{sec3} can be applied also to the pion case, replacing the non-perturbative inputs 
by the corresponding pion quantities, which we denote with the superscript ``$\pi$''. 
\begin{figure}[hbtp]
\centering
\hfill
\begin{minipage}{0.35\columnwidth}
\centering
\includegraphics[width=\columnwidth]{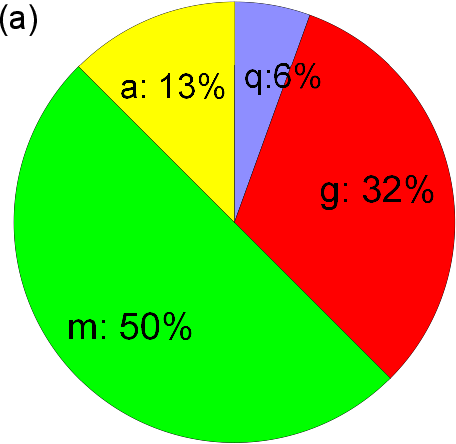}
\vspace{0.5cm}
\end{minipage}
\hfill
\begin{minipage}{0.6\columnwidth}
\centering
\includegraphics[width=\columnwidth]{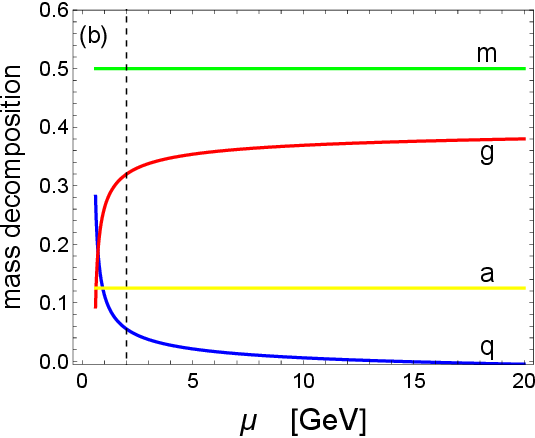}
\end{minipage}
\caption{Same as Fig.~\ref{piechart}, but for the pion case.}
\label{piechartpi}
\end{figure}
As explained in~\cite{Tanaka:2022wzy}, 
the global QCD analyses for the pion's PDFs are available at the NLO level, and the resultant value,
$A_q^\pi (0, \mu_0)=0.58(9)$ with $\mu_0=2$~GeV~\cite{Barry:2021osv}, has rather large ($\sim$ ten percent level) uncertainties (see also \cite{Barry:2018ort,Novikov:2020snp,ExtendedTwistedMass:2021rdx}),
while chiral perturbation theory gives the accurate value for $M_m^\pi=\frac{1}{2M^\pi}\langle \pi|\left[m \bar{\psi}
\psi \right]|\pi\rangle$ corresponding to (\ref{m3q}): $M_m^\pi  =  \frac{1}{2}M^\pi$, to the ${\cal O}(6\% )$ correction~\cite{Gasser:1980sb,Colangelo:2001sp}.
Using these input values, $A_r^\pi (0, \mu )$ and $\bar C^\pi_r(0,\mu )$ have been calculated as functions of $\mu$ by our NNLO formulas in the $\overline{\rm MS}$ scheme; e.g.,
$\bar C^\pi_q(0, \mu_0) |_{{n_f} = 3}=-\bar C^\pi_g(0, \mu_0) |_{{n_f} = 3}=-0.03(2)$
with $\mu_0=2$~GeV~\cite{Tanaka:2022wzy}.
These values allow us to evaluate the pion-mass decompositions corresponding to the four-term formulas (\ref{4term}) and (\ref{new4term}), as displayed in 
Figs.~\ref{piechartpi} and \ref{fig4}, 
respectively,
where the presented results have  uncertainties of $\sim$ ten percent level;
we recognize that, also in the pion case,
the points (i)-(iii) discussed in Sec.~\ref{sec2}
are resolved with (\ref{new4term}).

The comparison between Figs.~\ref{fig2} 
and~\ref{fig4} reveals that the twist-two contributions behave similarly, while
the proton and the pion are distinguished by the behaviors of the twist-four components, as driven by the large
quark-mass term $M_m^\pi$ for the pion, which reflects the Nambu-Goldstone boson nature.
Such a difference could not be observed if the last two terms of twist four in (\ref{new4term}) were not subdivided into the quark and gluon contributions, because $\frac{1}{4}{\cal M}_q+\frac{1}{4}{\cal M}_g=\frac{1}{4}M$;
on the other hand,
such a difference could be observed also when the ``D2 scheme''~\cite{Metz:2020vxd,Lorce:2021xku} is employed as in \cite{Liu:2021gco,Liu:2023cse} for the twist-four contributions in (\ref{new4term}), corresponding to making the formal replacements,
$\frac{1}{4}{\cal M}_q \to \frac{1}{4}M_m$ and $\frac{1}{4}{\cal M}_g \to M_a$ (see (\ref{ma})).
It is worth noting, as mentioned in Sec.~\ref{sec2}, that the D2 scheme is obtained by separating 
the traceless part and the trace part for each of the {\it bare} quark and gluon contributions~(\ref{tg}) in the
EMT, i.e., it corresponds to the subdivision of the trace part (twist-four part) into the quark and gluon contributions referring to the {\it bare} degrees of freedom.
On the other hand, the $\overline{\rm MS}$ renormalization is formulated referring to the renormalized quark and gluon degrees of freedom~\cite{Collins:1984xc},
and, in this paper,
we stick to the $\overline{\rm MS}$ scheme at all steps to treat the twist-four effects, as well as the twist-two contributions, for their quantitative evaluations.
\begin{figure}[hbtp]
\centering
\hfill
\begin{minipage}{0.35\columnwidth}
\centering
\includegraphics[width=\columnwidth]{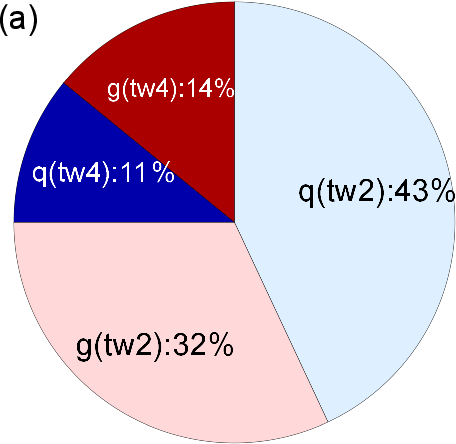}
\vspace{0.5cm}
\end{minipage}
\hfill
\begin{minipage}{0.6\columnwidth}
\centering
\includegraphics[width=\columnwidth]{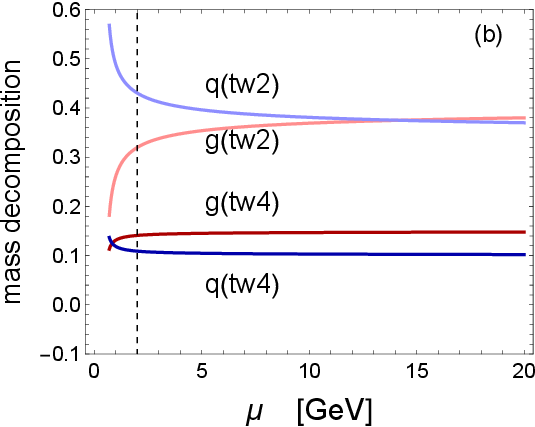}
\end{minipage}
\caption{Same as Fig.~\ref{fig2}, but for the pion case.}
\label{fig4}
\end{figure}

\section{Conclusions}
\label{sec5}

The twist-four contributions in Figs.~\ref{fig2},~\ref{fig4} have a very weak $\mu$ dependence, and the corresponding values multiplied by $4$ also represent the results for the quark/gluon decomposition~(\ref{calmqg0}) of the invariant mass of (\ref{m22})~\cite{Tanaka:2018nae,Tanaka:2025yez}.
This decomposition~(\ref{calmqg0}), composed only of the twist-four effects, coincides in total 
with the rest-mass decomposition~(\ref{new4term}) that also involves the twist-two effects;
this equality is 
guaranteed by the sum rule mentioned below (\ref{para}), i.e., $\bar{C}_q+ \bar C_g=0$; compare (\ref{new4term}) and (\ref{calmqg}).
Apparently, the different characteristics arising in those different types of decompositions are 
caused by the behavior of 
$\bar{C}_q(0, \mu)$ ($=-\bar{C}_g(0, \mu)$), 
which is actually associated with the twist-four 
three-parton correlation~\cite{Tanaka:2018wea,Tanaka:2018nae}.
Indeed, extending the treatment of the ``virial theorem'' in \cite{Lorce:2021xku} based on the dilatation transformation, we are able to obtain the exact relation,
\begin{align}
&\bar{C}_q(0, \mu)= \frac{1}{3M}\Bigl\langle \int {{d^3}} x x_\beta \left[\bar \psi(x) \gamma_\alpha\, g{F^{\alpha\beta}}(x) \psi (x) \right]\Bigr\rangle\ ,
\label{corrcorrq}
\\
&\bar{C}_g(0, \mu)= \frac{1}{3M}\Bigl\langle \int d^3 x {x_\alpha } \left[F_a^{\alpha\lambda}(x)D_{ab}^\beta F_{\beta \lambda }^b(x)\right]\Bigr\rangle\ ,
\label{corrcorrg}
\end{align}
corresponding to the forces between the quarks and gluons, 
and to those
among the gluons, the details of which will be discussed elsewhere.

Our new four-term mass-decomposition formula~(\ref{new4term}) may be regarded as the ``hybrid'' mass-decomposition formula with the combined use of 
both formulas in (\ref{m22}), the rest energy and the invariant mass: (\ref{new4term}) has been derived starting from the former, with its twist-four contribution being decomposed in terms of (\ref{calmqg0}) due to the latter.
It should be emphasized that, for such hybrid-type formulas strictly separating into the traceless part and the trace part,
(\ref{new4term}) gives a unique form when the MS-like schemes are thoroughly employed to handle the composite operators of twist four as well as twist two, for each of the gauge-invariant quark part and gluon part of the EMT that obeys the energy-momentum conservation of~(\ref{renomcontinue}).
Evaluating each four-term of (\ref{new4term}) in the NNLO QCD,
we have found that the first two terms of twist two represent the partonic motion inside a hadron, showing similar behaviors for the proton and the pion, while the latter two terms of twist four are induced by partonic correlations, showing quite different behaviors between the proton and the pion. 

Thus, the mass
of a hadron could be regarded as composed of the universal quark and gluon contributions due to their internal motions, combined with the correlation contributions due to their mutual interactions, (\ref{corrcorrq}), (\ref{corrcorrg}), emergent as non-perturbative QCD effects; the quantitative effects of the former are determined by the average momentum fractions of quarks/gluons as the moment of 
the corresponding PDFs, while those of the latter are controlled by the anomalous breaking of scale invariance. 

In conclusion, we have evaluated the mass-decomposition formulas of the proton and the pion, expressing them in terms of the quark/gluon gravitational form factors in QCD.
Using the recent NNLO QCD results for the forward values of the relevant gravitational form factors,
we have succeeded in obtaining the NNLO update of quantitative values for the proton-mass as well as pion-mass
decompositions proposed in the literature, with high accuracy, in particular, at
a few percent level for the proton. The NNLO RG evolution of those results has also been obtained over a range of renormalization scales $\lesssim 20$~GeV.
We have also performed similar calculations using our new four-term mass-decomposition formula~(\ref{new4term}), 
which is obtained by making the traceless/trace decomposition mathematically precise for each of the quark and gluon contributions in the MS-like schemes.
The results for the proton and pion masses have demonstrated the advantages of this new decomposition, indicating that it provides clarification of the physical effects inside the hadron,
into those from the motions of the partons, and from the correlations due to non-perturbative interactions among the partons;
the partonic motions have similar effects for the proton and the pion, while the correlations yield quite different effects, distinguishing between the proton and the pion.
Those physical pictures emerging in the hadron mass decompositions based on the QCD EMT deserve further exploration,
particularly in the context of confinement, and should also be examined in other aspects of hadron structure. 

\begin{acknowledgments}
The author thanks Yoshitaka Hatta for useful discussions concerning the renormalization schemes, and Hiroyuki Kawamura and Yuichiro Kiyo for insightful discussions.
This work was supported by JSPS KAKENHI Grant Numbers JP24K07055 and JP23K03419.
\end{acknowledgments}



\bibliography{pion}

\end{document}